\documentclass[prb,showpacs,twocolumn,groupedaddress,showemail]{revtex4}
\usepackage{amsmath}
\usepackage{graphicx}
\begin{document}
\title{First-principles study of a single-molecule
magnet Mn$_{12}$ monolayer on the Au(111) surface}
\author{Salvador Barraza-Lopez}\email{salva@vt.edu}
\author{Michael C. Avery} 
\author{Kyungwha Park}\email{kyungwha@vt.edu}
\affiliation{Department of Physics, Virginia Polytechnic Institute 
and State University, Blacksburg, VA 24061.}
\begin{abstract}
The electronic structure of a monolayer of single-molecule magnets 
Mn$_{12}$ on a Au(111) surface is studied using spin-polarized 
density-functional theory. The Mn$_{12}$ molecules are oriented
such that the magnetic easy axis is normal to the surface, and 
the terminating ligands in the Mn$_{12}$ are replaced by thiol 
groups (-SH) where the H atoms are lost upon adsorption onto the
surface. This sulfur-terminated Mn$_{12}$ molecule has a total magnetic 
moment of 18~$\mu_B$ in the ground state, in contrast to 20$\mu_B$
for the standard Mn$_{12}$. The Mn$_{12}$ molecular orbitals broaden
due to the interaction of the molecule with the gold surface and
the broadening is of the order of 0.1~eV. It is an order of magnitude less
than the single-electron charging energy of the molecule so the molecule
is weakly bonded to the surface. Only electrons with majority spin
can be transferred from the surface to the sulfur-terminated Mn$_{12}$
since the gold Fermi level is well above the majority lowest 
unoccupied molecular orbital (LUMO) but below the minority LUMO.
The amount of the charge transfer is calculated to be 1.23~electrons from 
a one-dimensional charge density difference between the sulfur-terminated 
Mn$_{12}$ on the gold surface and the sulfur-terminated Mn$_{12}$, dominated by the
tail in the electronic distribution of the gold surface. A calculation of a level shift
upon charging provides 0.28 electrons being transferred. The majority of the 
charge transfer occurs at the sulfur, carbon, 
and oxygen atoms close to the surface. The total magnetic moment 
also changes from 18~$\mu_B$ to 20~$\mu_B$, which is due to rearrangements 
of the magnetic moments on the sulfur atoms and Mn atoms upon 
adsorption onto the surface. The magnetic anisotropy 
barrier is computed including spin-orbit interaction self-consistently 
in density-functional theory. The barrier for the Mn$_{12}$ on the gold 
surface decreases by 6~K in comparison to that for an isolated 
Mn$_{12}$ molecule.
 
\end{abstract}
\date{\today}

\maketitle

\section{Introduction}

A nanoscale single-molecule magnet (SMM) comprises a few transition metal ions 
interacting through organic and/or inorganic ligands instead of direct exchange
interactions. To reverse the magnetic moment of a SMM, a large energy barrier 
needs to be overcome. Experiments on bulk forms 
of SMMs exhibited quantum tunneling between different directions of magnetic
moments, \cite{FRIE96,THOM96} quantum interference between spin paths,\cite{WERN99} 
and long spin dephasing time $T_2$.\cite{ARDA07} These properties of SMMs 
propelled interest in utilizing SMMs as information storage 
devices,\cite{JOAC00} spin-based devices,\cite{TIMM06} or materials for quantum 
computation.\cite{LEUE01} A great amount of experiments were carried out
on deposition of SMMs Mn$_{12}$ or its derivatives on gold \cite{STEC04,ZOBB05,NAIT04,NAIT05} 
and silicon \cite{FLEU05,MART07,SALM07} 
surfaces or on bridging them between gold electrodes \cite{HEER06,Jode6,delBarco07}. 
Among thousands of synthesized SMMs, [Mn$_{12}$O$_{12}$(CH$_3$COO)$_{16}$(H$_2$O)$_4$] 
(referred to as Mn$_{12}$) was widely studied due to its
large magnetic anisotropy barrier (MAB) or magnetization 
reversal barrier of 65~K.\cite{BARR97} 
Largely, Mn$_{12}$ molecules were deposited onto a surface or bridged 
between electrodes in two different manners: (i) through attractive 
van der Waals forces between the surface and Mn$_{12}$ without 
surface-binding ligands or (ii) via ligand exchange 
with the Mn$_{12}$ molecules. In the latter case, for example, carboxylate 
terminated alkanethiolates, C$_n$H$_{2n+2}$S, would be surface-binding 
ligands between the molecules and the surface,\cite{STEC04} or 16 acetate
(O$_2$C-CH$_3$ or Ac) ligands within Mn$_{12}$ could be replaced by ligands 
O$_2$C-C$_6$H$_4$-SAc that create a direct strong bond to gold.\cite{HEER06} 

Scanning tunneling microscope (STM) images \cite{ZOBB05,NAIT04,VOSS07} 
and atomic force microscope (AFM) images \cite{FLEU05} on the monolayers of Mn$_{12}$ 
molecules revealed that the Mn$_{12}$ molecules in the monolayers are 
individually distinguishable rather than aggregated on the surface. 
Photoemission spectroscopy experiments on the monolayers of Mn$_{12}$ derivatives 
showed that the Mn 3$d$ partial density of states in valence bands 
for the monolayers is comparable to that for bulk Mn$_{12}$. \cite{DELPE06,VOSS07-PRB} 
A low concentration of molecules on surfaces makes it challenging to 
accurately measure the magnetic properties of a single monolayer of Mn$_{12}$. 
A recent magnetic measurement on thick monolayers of Mn$_{12}$ molecules 
exhibited qualitatively different magnetic properties from bulk Mn$_{12}$. 
\cite{NAIT05} Very recently, local magnetic properties of a Mn$_{12}$ monolayer 
were measured using depth-controlled $\beta$-detected nuclear magnetic
resonance (NMR). This measurement also suggested that the magnetic properties 
differed from those of bulk Mn$_{12}$. \cite{SALM07} 
In all of the experimental systems discussed little was known about the 
interface between the Mn$_{12}$ molecules and the surface or electrodes. 
For example, the following questions have not been answered: (i) how the 
Mn$_{12}$ molecules are oriented at the interface, (ii) whether the Mn$_{12}$ 
molecules remain chemically intact at the interface, and (iii) how strongly 
the molecules are coupled to a surface or electrodes. Therefore, it is still 
a controversy whether the electronic and magnetic properties of SMMs change 
due to the interaction with a surface.

To investigate mainly electronic transport and spin filtering through SMMs,
theoretical models \cite{KIM04,ELST06,ROME06-2,ROME06-3,LEUE06} 
were proposed based on the assumption that individual SMMs were not strongly 
coupled to electrodes, in other words, the molecular orbitals remain  
sharp despite their interactions with metal electrodes. Then the Anderson 
Hamiltonian and an effective spin Hamiltonian were used
with a priori microscopic parameter values that need to be determined from 
atomic-scale simulations. To the best of our knowledge, SMMs deposited 
on a surface or bridged between electrodes, have not been yet studied using
first-principles methods. We investigate, within density-functional theory 
(DFT), the effect of the interface on the electronic and magnetic properties 
of Mn$_{12}$ molecules on a gold surface, given a particular orientation 
of Mn$_{12}$ molecules relative to the surface. More specifically we obtain 
the strength of coupling of the molecules to the surface and other parameter 
values that could be used in a model Hamiltonian.

\begin{figure}
\includegraphics[angle=0, width=0.45\textwidth]{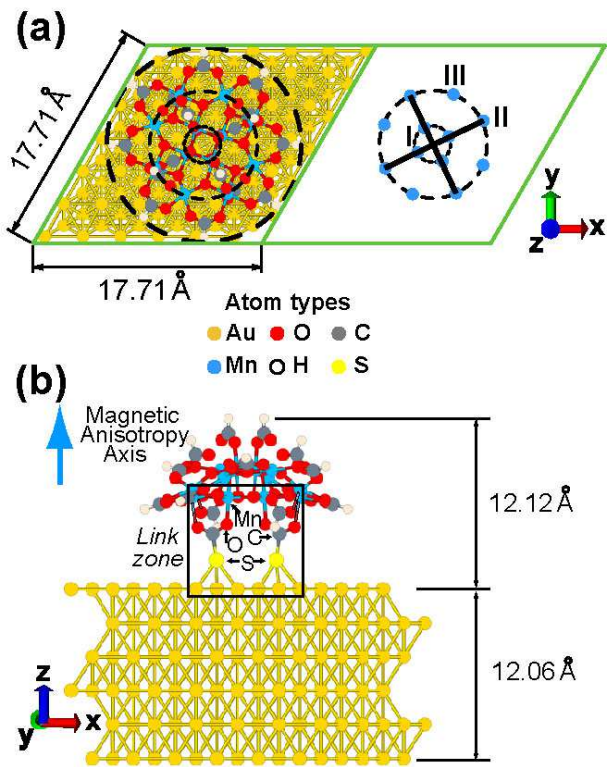}
\caption{\label{fig:1}(Color online) Schematic (a) top and (b) side views of
a Mn$_{12}$ molecule adsorbed onto a gold slab via thiol groups. The three
different Mn ions (regarding symmetry) are denoted as I, II, and III in (a).
The side view of the six gold monolayers with 36 surface atoms per monolayer 
are shown in (b). In supercell calculations, a vacuum layer of 10~\AA{} 
is placed above the Mn$_{12}$ molecule.}
\end{figure}

We consider a monolayer of Mn$_{12}$ molecules adsorbed onto a Au(111) surface 
through thiol groups as shown in Fig.~\ref{fig:1}. Here we use the shortest 
chemical link between the Mn$_{12}$ molecules and the gold surface. 
Although this link may seem too short to create stable Mn$_{12}$ monolayers 
in experiments, it has advantages for theoretical studies. The link 
would considerably reduce computational cost by requiring a smaller unit cell
without altering the physical and chemical properties of the system.
If the molecules can be strongly coupled to the surface, this
link would be a good candidate. Our DFT calculations show that the 
Mn$_{12}$ molecular orbitals moderately broaden for the short link due to
the interaction between the molecules and the surface.
Electronic charge is transferred from the surface to the molecules and 
the total magnetic moment is also modified upon adsorption onto the surface. 
Our model and methods for studying a Mn$_{12}$ monolayer on the gold surface 
are presented in Sec.~II. The electronic structure and magnetic properties on 
the system are discussed in Sec.~III. The conclusion follows.

\section{Model and methods}
Spin-polarized DFT calculations are performed with plane waves as basis sets,
within the Perdew-Burke-Ernzerhof (PBE) generalized-gradient approximation (GGA),
\cite{PERD98} using the Vienna Ab-initio Simulation Package (VASP).\cite{VASP1,VASP2} 
Projector-augmented-wave (PAW) pseudopotentials \cite{Blochl,PAW} are employed 
to take into account spin-orbit interaction (SOI) that induces large magnetic
anisotropy in a Mn$_{12}$ molecule. Valence electrons considered in the PAW 
pseudopotentials for each atom are shown in Table \ref{ta:table1}. 
$5d^{10}6s^1$ orbitals are treated as valence states for Au, and $3p^6 4s^2 3d^5$ 
orbitals are used for Mn. Hard pseudopotentials are used for C and O.
All DFT calculations in this study are performed self-consistently 
using VASP (unless stated otherwise) 
until the total energy converges down to 1.0$\times{}10^{-5}$ eV. 
To simulate a monolayer of Mn$_{12}$ molecules on a Au(111) surface (Fig.~\ref{fig:1}), 
we consider a monoclinic unit cell of 17.71$\times$17.71$\times$34.00~\AA$^3${} 
in which a simplified form of a Mn$_{12}$ molecule (explained specifically
in Sec.II.B) is attached to a gold slab with six monolayers 
via thiol groups, and a vacuum layer of 10~\AA{} is added above the 
Mn$_{12}$ molecule. In this geometry the magnetic easy axis of the 
Mn$_{12}$ molecules ($z$ axis) is oriented normal to the surface. 
We first optimize a gold slab and a Mn$_{12}$ molecule geometry, 
separately, and then combine them to create what we call the 
`whole structure' illustrated in Fig.~\ref{fig:1}. 

\subsection{Gold slab}
A DFT calculation is performed on bulk gold and convergence 
of the total energy is checked as a function of the number 
of $k$-points, fast Fourier transform (FFT) mesh, an energy 
cutoff for plane waves, and a cutoff in augmentation charges. 
With an energy cutoff of 260~eV and a cutoff in augmentation
charges of 357~eV, the calculated equilibrium lattice constant 
is 4.175~\AA{} that differs from the experimental value\cite{Khein95} 
by 2.8\%. In constructing the gold slab, in-plane separations
among gold atoms are set to be 2.952 \AA{}, obtained from the
equilibrium geometry of bulk gold. With the in-plane separations 
fixed, the distance between monolayers is determined by atomic 
force relaxation until the magnitude of all force components 
is less than 0.01~eV/\AA.{} 
To fully cover a Mn$_{12}$ molecule on a Au(111) surface, 
a 6$\times$6$\times$1 real-space supercell (36 surface Au atoms 
per monolayer) is used. The surface area covered by the molecule is 
stressed by the dotted circle in Fig.~\ref{fig:1}(a). In addition, 
a gold slab should be thick enough to completely screen the Mn$_{12}$ 
molecule including thiol groups with height of 12.12~\AA.{} This 
condition is satisfied with six gold monolayers 12.06~\AA{} high. 
In the whole structure we include the optimized geometry for 
a six-monolayer gold slab with 36 surface atoms 
per monolayer that contains a total of 216 gold atoms. 

\subsection{Isolated Mn$_{12}$: geometries 1 and 2}

\begin{figure}
\includegraphics[angle=0, width=0.45\textwidth]{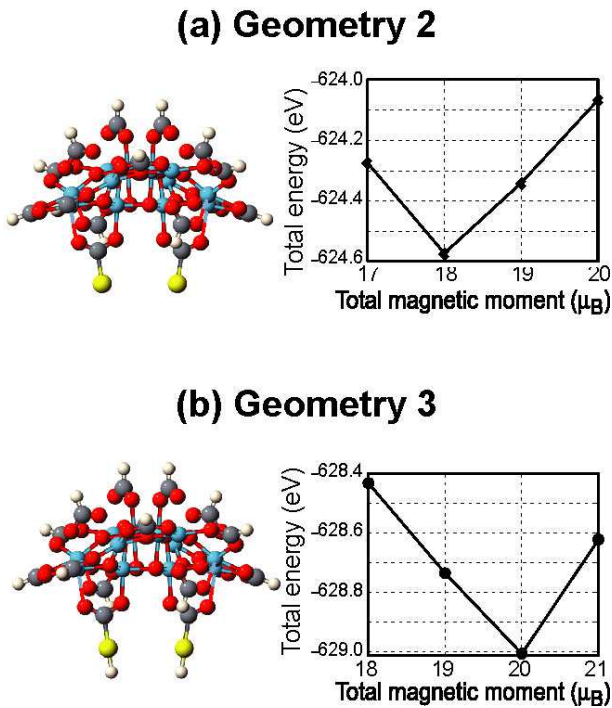}
\caption{\label{fig:2}(Color online) (a) Geometry of the sulfur-terminated  
Mn$_{12}$, geometry {\bf 2}, and its total energy as a function of
total magnetic moment. (b) Geometry of the thiol-terminated Mn$_{12}$, geometry
{\bf 3}, and its total energy as a function of total magnetic moment.}
\end{figure}

To reduce computational cost, a simplified form of a Mn$_{12}$ molecule, 
[Mn$_{12}$O$_{12}$(HCOO)$_{16}$] (denoted geometry {\bf 1}), is used 
in our study. The structure of geometry {\bf 1} is the same as what is
shown in Fig.\ref{fig:2}(a) except that the S atoms are replaced by H.
The simplified Mn$_{12}$ differs from the synthesized 
Mn$_{12}$ molecule in that the 16 acetate ligands were replaced by 16 
formates (HCOO) and the four water molecules were removed because 
they may be lost via ligand-exchange during adsorption onto a surface.\cite{STEC04} 
Geometry {\bf 1} is placed in a unit cell of 
$25.0 \times 25.0 \times 20.0$~\AA$^3$, where the magnetic easy axis
is along the $z$ axis. Then geometry {\bf 1} is relaxed with a fixed
total magnetic moment of 20~$\mu_B$ (Bohr magneton) and an energy cutoff 
of 600~eV until the magnitude of all force components becomes less than 
0.08~eV/\AA.{} Using S$_4$ symmetry of geometry {\bf 1}, the twelve Mn ions 
are categorized into three classes marked as I, II, and III in 
Fig.~\ref{fig:1}(a). The four inner Mn ions (Mn$^{4+}$) belong to symmetry 
class I, while the two sets of the four Mn ions (Mn$^{3+}$) in the outer 
ring belong to symmetry classes II and III, respectively. 
The spin moments of the Mn$^{4+}$ ions are antiparallel to those of
the Mn$^{3+}$ ions in the ground state. Our noncollinear calculation
shows that a collinear spin approximation is good for geometry 
{\bf 1}. The simplification made for the Mn$_{12}$ molecule does not 
significantly change the electronic and magnetic properties of the
molecule. The gap between the majority highest occupied molecular 
orbital (HOMO) and the majority lowest unoccupied molecular orbital 
(LUMO) is calculated to be 0.24~eV, while the minority HOMO-LUMO gap
is 1.95~eV. 
%
%

\begin{table}
\caption{\label{ta:table1} Valence electrons, energy cutoffs, and atomic sphere 
radii for the atoms where the PAW pseudopotentials are employed. The given atomic 
radii are used to calculate atomically resolved magnetic moments shown in 
Table~\ref{ta:table2}.}
\begin{ruledtabular}
\begin{tabular}{cccc}
Atomic          & Valence Electron  & Suggested Energy         & Atomic Sphere \\
Species         & Configuration     & Cutoff Range (eV)  & Radii (\AA)   \\
\hline
Au              & $5d^{10}6s^1$            &   170-230      & 1.50 \\
Mn              & $3p^64s^23d^5$            &   202-270      & 1.32 \\
O               & $2s^22p^4$            &   500-700      & 0.74 \\
C               & $2s^22p^2$            &   500-700      & 0.86 \\
S               & $3s^23p^4$            &   302-402      & 0.95 \\ 
H               & $1s^1$               &   200-250      & 0.37 
\end{tabular}
\end{ruledtabular}
\end{table}

\begin{table*}
\caption{\label{ta:table2} Atomically resolved magnetic moments for 
geometry {\bf 2} and atomically resolved difference of the magnetic moments 
between the whole structure 
and geometry {\bf 2} in units of Bohr magneton. The positive sign in the magnetic
moment change denotes an increase in the magnetic moment upon adsorption. The three 
different Mn ions in terms of symmetry, Mn(I), Mn(II), and Mn(III), are depicted 
in Fig.~\ref{fig:1}(a). Significant increases in the magnetic moments occur at 
the atoms whose magnetic moments are bold-faced.}
\begin{ruledtabular}
\begin{tabular}{l|c|c}
Atomic species & Initial magnetic moment & Change of magnetic moment upon adsorption    \\
\hline
Mn (I)          &  $-$2.621,$-$2.620,$-$2.573,$-$2.571      & {\bf +0.019,+0.021},+0.002,+0.001\\
Mn (II)         &    3.453,3.461,3.530,3.516            & {\bf +0.063,+0.060},+0.004,+0.004\\
Mn (III)        &    3.541,3.543,3.527,3.534            & +0.002,+0.003,+0.010,+0.010\\
\hline
O closest to Mn and S   & $-$0.011,$-$0.011,$-$0.028,$-$0.027   & +0.015,+0.014,{\bf +0.036,+0.035}\\ 
O (total)       & 0.328                 & +0.105\\
\hline
C closest to S      & 0.007,0.007               &$-$0.008,$-$0.008\\
C (total)       &0.138                  &$-$0.016\\
\hline
S           &$-$0.236, $-$0.235         &{\bf +0.237,+0.238}
\end{tabular}
\end{ruledtabular}
\end{table*}

\begin{figure}
\includegraphics[angle=0, width=0.45\textwidth]{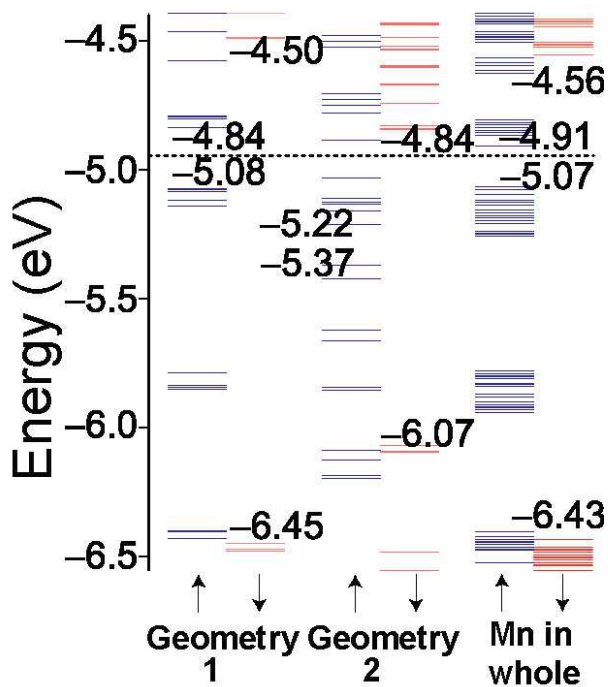}
\caption{\label{fig:3}(Color online) Molecular orbitals in the vicinity of the 
gold Fermi level (dashed line) for geometries {\bf 1} and {\bf 2} and the whole 
structure. The orbital energies are written for the majority and minority HOMO and 
LUMO. The orbitals shown for the whole structure are projected ones onto all the 
Mn atoms only.}
\end{figure}

To attach the Mn$_{12}$ molecules onto a gold surface, the H 
atoms in the two formates in geometry {\bf 1} are exchanged with 
surface-binding ligands, SH. 
Then the H atoms in SH are removed for the S atoms to 
be directly bonded to the gold surface. This determines the 
geometry {\bf 2}, as shown in Fig.~\ref{fig:2}(a). The distance 
between the S atom and each bonding C atom is 
1.88~\AA,~which was obtained from geometry optimization of 
Mn$_{12}$-alkane-S-Au$_{13}$. The Fermi level of the gold slab 
as well as molecular orbitals in its vicinity including HOMO and 
LUMO are depicted in Fig.~\ref{fig:3} for geometries {\bf 1} and {\bf 2} 
and the whole structure. In geometry {\bf 2} the S atoms are highly reactive and their 
presence reduces both the majority and minority HOMO-LUMO gaps 
compared to geometry {\bf 1}. Due to the lost H atoms, the HOMO and 
LUMO of geometry {\bf 2} are shifted downward compared to geometry 
{\bf 1} and the total magnetic moment of geometry {\bf 2} decreases 
to 18~$\mu_B$. The calculated total energy of geometry {\bf 2} as 
a function of total magnetic moment is shown in 
Fig.~\ref{fig:2}(a). Magnetic moments for individual atoms are 
calculated over atom-centered spheres with radii given in Table~I. 
See Table~II. The S atoms bear substantial magnetic moments 
of 0.24~$\mu_B$ parallel to the inner Mn(I) ions. The magnetic moments 
of the Mn(I) and Mn(II) ions (denoted bold in Table~II) connected to the 
S atoms via the O and C atoms are considerably smaller 
than those for geometry {\bf 1}.

\subsection{Whole structure: geometry 2 on Au(111)}

Top and side views of the whole structure used in our DFT calculation 
are shown in Fig.~\ref{fig:1}. 
Each sulfur atom is bonded to the closest hollow (face-centered cubic) 
site on the gold slab which is the most favorable configuration. 
Attaching a sulfur atom at a bridge site requires a 0.30~eV higher energy 
than at a hollow site. 
This result agrees with a calculation on alkane thiols on a gold surface, 
for which a hollow site was favored by 0.40~eV.\cite{Rappe01} A top site on 
a gold slab is the most expensive energetically. The distance from the sulfur 
atoms to the gold slab is optimized such that both sulfur atoms are as close as 
possible to hollow sites. 
The distances between the sulfur and the gold atoms range between 2.52 
and 2.74~\AA{}, where the longest bond is due to a slight mismatch of placing 
one of the sulfur atoms at the closest hollow site on the gold slab.

In the whole structure the closest distance between hydrogen atoms on periodic 
images of the Mn$_{12}$ (geometry {\bf 2}) is 3.35 \AA.{}
The large intermolecular separation prevents different Mn$_{12}$ molecules 
from interacting with one another except for dipolar interactions. In the present
study the dipolar interactions are not considered. The whole structure has a total of 
304 atoms and 2886 valence electrons. Due to the presence of Mn atoms, an
energy cutoff of 600~eV for plane waves and a cutoff in the augmentation 
charges of 800~eV are used. For the whole structure, we sample 
2$\times$2$\times$1 $k$-points, including the $\Gamma$ point. Notice that
we used a 6$\times$6$\times$1 real-space supercell per gold monolayer.
Thus, on the plane parallel to the gold surface, our $k$-point sampling
is equivalent to having (6$\cdot$2)$\times$(6$\cdot$2)=12$\times$12 $k$-points 
in a 1$\times$1$\times$1 real-space gold cell. A further
increase in the number of $k$-points does not affect our results.

In a self-consistent calculation on the whole structure,
the first-order Methfessel-Paxton scheme is used with the smearing 
parameter $\sigma$=6.0~meV. This choice of $\sigma$ is necessary 
to resolve individual energy levels in projected density of states (PDOS) 
and to compute the MAB caused by the SOI 
in the Mn$_{12}$ molecule. 
The total magnetic moment is set to 20~$\mu_B$ and the discussion
in Sec.~III suggests that 20~$\mu_B$ is the ground-state total
magnetic moment for the whole structure. The combination of high energy 
cutoffs and a large number of valence electrons involved indicates the 
numerically expensive nature of the performed calculations so 
we do not further relax the whole structure. A comprehensive analysis 
of the forces exerted on the structure is provided.
The average over \emph{absolute values} of all of the 
$x$, $y$, and $z$ force components is (0.03,0.03,0.05) eV/\AA~ 
with standard deviation of (0.08,0.07,0.08) eV/\AA.~Out of a total of 
304 atoms, forces on only 25 atoms are greater than 0.15 eV/\AA{} 
and the large forces occur near the sulfur atoms. The $x$ and $y$ 
components of the forces on the gold atoms bonded to the sulfur 
atoms are 0.60 eV/\AA{} and the $z$ force components are 0.20 eV/\AA{} 
with alternating signs. The forces on nearest 
neighbors to those gold atoms are 0.20 eV/\AA~along the $x$ and $y$ axes. 
The force on one of the sulfur atoms has components (0.57,0.29,0.13)~eV/\AA~
because of the mismatch of placing the sulfur 
atom at a hollow site of the gold slab. The forces onto the two carbon 
atoms bonded to the sulfur atoms are 0.39 eV/\AA{} along the $z$ axis.
The oxygen atoms bonded to those carbon atoms have $z$ force
components that fall between $-$0.66 and $-$0.52 eV/\AA.{} The Mn
ions connected to the S via C and O atoms (two of the Mn(I) 
and two of the Mn(II) ions) have $z$ force components of
0.24~eV/\AA.~ 
The analysis of the forces indicates that relaxation of the whole structure 
may possibly break the symmetries of the Mn(I) and Mn(II) ions, and could 
further modify its magnetic properties. 

\section{Results and Discussion}
The electronic structures of the six gold monolayers, geometries {\bf 1} and 
{\bf 2}, and the whole structure are calculated and discussed using the PDOS onto 
particular orbitals of specific atoms. A comparison of the PDOS for the whole 
structure (Fig.~\ref{fig:6}) with those for geometry {\bf 2} (Fig.~\ref{fig:5}), 
computed under identical conditions, reveals that the molecule of geometry {\bf 2} 
is weakly coupled to the gold surface for the given link, and that the molecular 
orbitals broaden by at most an order of 0.1~eV. The Fermi level of the gold slab 
is well above the majority LUMO but below the minority LUMO of geometry {\bf 2}
(Fig.~\ref{fig:3}). As a result, only electrons with majority spin can be 
transferred from the gold slab to the molecule. The ground-state total magnetic 
moment can also change upon adsorption on the surface due to strong bonds formed 
between the sulfur atoms and the gold slab.  
The amount of the charge transfer is quantified from 
two methods: (i) a one-dimensional electronic charge density difference between 
the whole structure and geometry {\bf 2}, and (ii) a shift in 
the lowest-lying molecular orbital as a function of extra partial charges. 
The change in the total spin magnetic moment is obtained from a
one-dimensional spin density difference. 
The effect of the charge transfer and spin magnetic moment change on the MAB
is discussed from self-consistent SOI calculations.

\subsection{Projected densities of states (PDOS)}

\begin{figure}
\includegraphics[angle=0, width=0.47\textwidth]{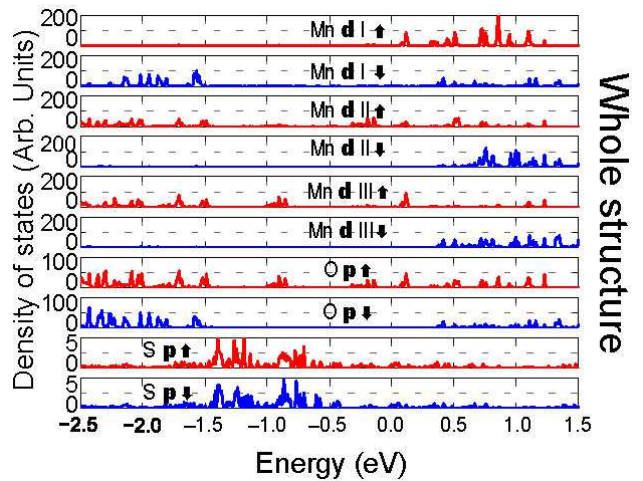}
\caption{\label{fig:6}(Color online) Projected densities of states onto
the majority and minority Mn $d$, O $p$, and S $p$ orbitals for the
whole structure: geometry {\bf 2} on Au(111). The zero in the horizontal
scale denotes the Fermi level.}
\end{figure}

\begin{figure}
\includegraphics[angle=0, width=0.47\textwidth]{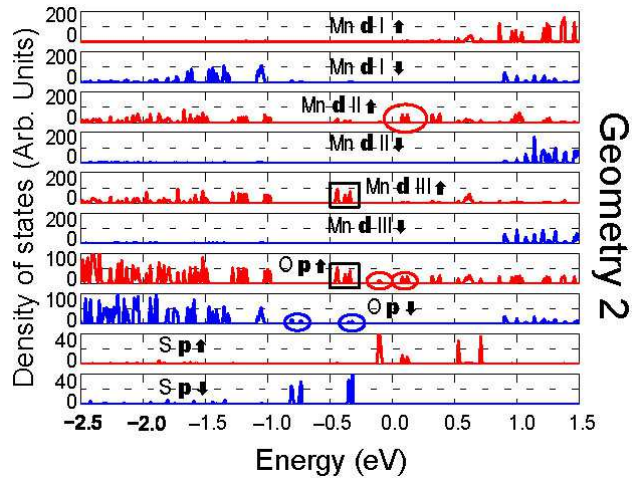}
\caption{\label{fig:5}(Color online) Projected densities of states onto
the majority and minority Mn $d$, O $p$, and S $p$ orbitals for 
geometry {\bf 2}. The zero in the horizontal scale denotes the midpoint
between the HOMO and LUMO energies. The ovals and squares correspond to 
the Mn $d$ and O $p$ orbitals hybridized with the S $p$ orbitals. 
Notice that the vertical scale for the S orbitals here differs from
that in Fig.\ref{fig:6}.}
\end{figure}

\begin{figure}
\includegraphics[angle=0, width=0.47\textwidth]{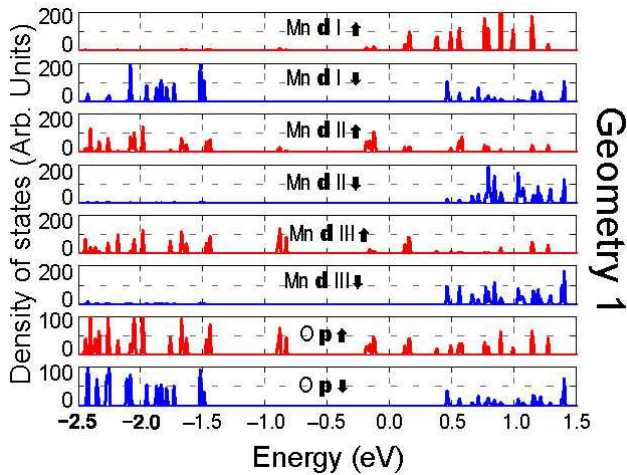}
\caption{\label{fig:4}(Color online) Projected densities of states onto
the majority and minority Mn $d$ and O $p$ orbitals for geometry {\bf 1}.
The zero in the horizontal scale denotes the midpoint
between the HOMO and LUMO energies.}
\end{figure}

The PDOS shown in Figs.~\ref{fig:6}, \ref{fig:5} and \ref{fig:4} are 
generated with 100 data points/eV and smearing parameter $\sigma$=6~meV 
without the SOI, over atom-centered spheres with the radii given in 
Table~\ref{ta:table1}. The horizontal scale in 
Figs.~\ref{fig:5} and \ref{fig:4} is shifted by 0.34~eV and 3.5~meV,
respectively, such that the zero in the horizontal scale corresponds 
to the midpoint of the HOMO and LUMO energies.
The PDOS onto the gold $s$ and $d$ orbitals (not shown) indicate a large 
density of states within 0.5~eV below and above the gold Fermi level.
To directly compare the PDOS for the whole structure with those for 
geometries {\bf 1} and {\bf 2}, \emph{exactly the same} 
parameter values are used for the three systems, such as an identical unit cell, 
$k$-point sampling, energy cutoff, cutoff in the augmentation charges, 
PDOS sampling parameters, smearing parameter, and total energy convergence 
tolerance. For geometry {\bf 2} the HOMO and orbitals right below 
the HOMO are mostly from the Mn(III) ions, neighboring O anions, and S atoms, while the 
LUMO and orbitals right above the LUMO are from the Mn(II) ions, 
neighboring O anions, and the S atoms. As highlighted by ovals 
and squares in Fig.~\ref{fig:5}, 
the S $p$ orbitals are hybridized with the Mn $d$ and O $p$ orbitals. 
Within 0.5~eV below the midpoint of the HOMO and LUMO energies, the spin 
moments of the S atoms are antiparallel to those for the Mn(III) and O atoms. 
For geometry {\bf 1} the HOMO and orbitals right below
the HOMO are mainly from the Mn(II) ions and O anions, while the LUMO and 
orbitals right above the LUMO are from the Mn(I) and Mn(III) ions and O 
anions (Fig.~\ref{fig:4}). These main differences of the PDOS between geometries 
{\bf 1} and {\bf 2} are linked to the difference of the total ground-state magnetic 
moment between them.

For the whole structure the PDOS (Fig.~\ref{fig:6}) onto the Mn $d$ and O $p$ 
orbitals noticeably differ from those for geometry {\bf 2} because of the 
difference in the total magnetic moment between them.
Although the PDOS for the whole structure are similar to those for geometry
{\bf 1}, the peak heights in the former are reduced and the orbitals broaden 
compared to the latter. The broadening is of an order of 0.1~eV 
(Figs.~\ref{fig:3}, \ref{fig:6}, and \ref{fig:4}). 
The single-electron charging energy for geometry {\bf 2} is calculated from the energy 
difference between a neutral molecule and one with an extra-electron added. When one 
extra electron is added to geometry {\bf 2}, the total ground-state magnetic 
moment increases to 19~$\mu_B$. Considering the change in the total magnetic
moment, we obtain the charging energy of 3.7~eV. 
This energy is one order of magnitude greater than the estimated broadening of 
the molecular orbitals so the molecule is weakly bonded to the surface.

\subsection{Charge transfer}

\begin{figure}
\includegraphics[angle=0, width=0.45\textwidth]{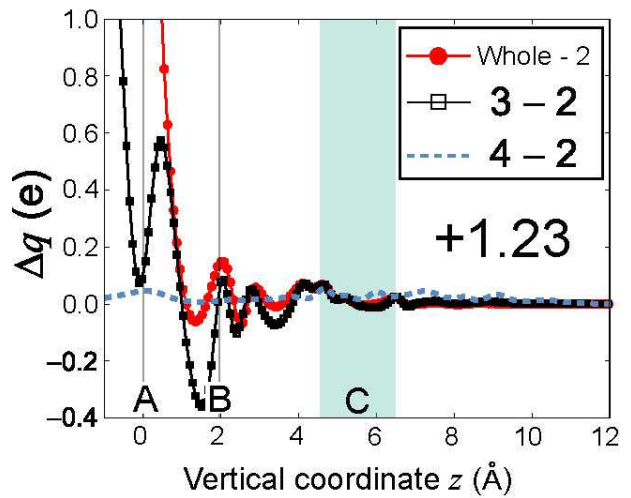}
\caption{\label{fig:7}(Color online) {\it Red filled circles}: 
Charge density difference between the whole structure and geometry {\bf 2} 
vs the $z$ coordinate. {\bf A}, {\bf B}, and {\bf C} denote the locations 
of the S atoms (at $z$=0.09~\AA),{} the bonding C atoms, and the Mn ions
illustrated in the inset of Fig.~\ref{fig:9}. Upon integration from $z=0.09$
($z=-0.93$) to $z$=12.00~\AA,{} a total of 1.23 (23.15) electrons is 
transferred from the surface to the molecule.
{\it Black empty squares}: Charge density difference between geometries 
{\bf 3} and {\bf 2}. {\it Blue dashed curve}: Charge density difference 
between a molecule with extra 0.3 electrons added to geometry {\bf 2} 
(denoted geometry {\bf 4}) and a neutral molecule with geometry {\bf 2}.}
\end{figure}

The charge transfer between the surface and the molecule of geometry {\bf 2} 
is computed from a one-dimensional electronic charge density difference 
between the whole structure and geometry {\bf 2} along the $z$ axis. 
The positive sign in the charge difference means an electronic charge is transferred 
from the gold surface to the molecule. The charge difference shown as the 
curve with filled circles in Fig.\ref{fig:7}, increases abruptly towards the sulfur 
atoms (vertical line {\bf A}) due to the electronic contribution of the 36 gold surface atoms 
as well as the sulfur atoms. (Each gold atoms has 11 valence electrons.) 
The deep penetration of the charge of the gold atoms beyond the sulfur atoms 
makes it difficult to precisely calculate the amount of the charge transfer. 
Including the long tail of the charge of the gold surface atoms, a charge of 
1.23 electrons (23.15 electrons) is transferred from the surface to the molecule
when the integration is performed from the sulfur atoms, $z=0.09$ (the mid-distance
between the first gold layer and sulfur atoms, $z=-0.93$)~ to $z$=12.00~\AA~ 
(Fig.~\ref{fig:7}). 
It is found that the transferred charge is mostly \emph{localized} in 
the sulfur and bonding carbon and oxygen atoms, while little charge is 
transferred to the Mn ions.
A similar trend is shown for the charge density difference between 
geometry {\bf 2} and the structure where two H atoms are bonded to 
the S atoms [denoted geometry {\bf 3}, Fig.~\ref{fig:3}(b)].
In this case, the charge difference obtained from integration
from $z=0.09$ to $z=12.00$~\AA,{} is 0.17 electrons. Notice that these two
sets of the charge density difference match well above vertical line
{\bf B} in Fig.~\ref{fig:7}, indicating that the electronic environment 
around the molecular magnetic core is very similar in the whole structure and 
geometry {\bf 3}. To further stress this point, we calculate the charge 
density difference between a neutral molecule with geometry {\bf 2} 
and the same molecule with 0.3 free electrons added. The extra free 
electrons are uniformly distributed over all atoms, as shown in Fig.~\ref{fig:7}.

\begin{figure}
\includegraphics[angle=0, width=0.45\textwidth]{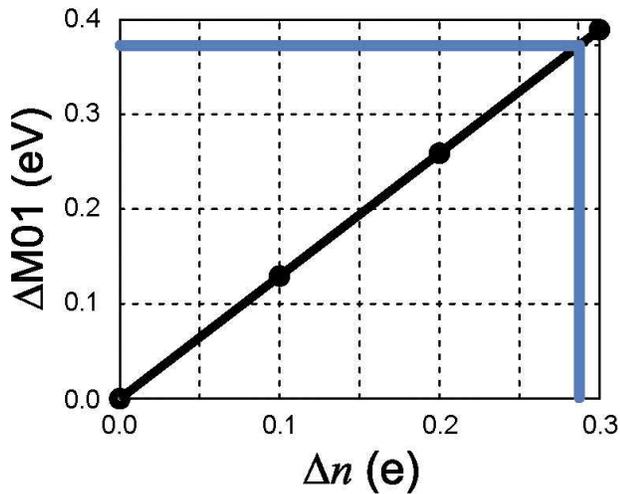}
\caption{\label{fig:8}(Color online) 
Shift of the lowest-lying molecular orbital (MO1) energy for geometry {\bf 2},
$\Delta$MO1, as a function of extra partial electronic charge, $\Delta n$. 
The thick horizontal line indicates the energy difference of the MO1 between 
the whole structure and geometry {\bf 2}.}
\end{figure}

The charge difference is also calculated from the change of the lowest-lying 
occupied molecular orbital (MO1) of geometry {\bf 2} as a function of extra 
partial electrons added to the molecule, following the procedure described
in Ref.~\onlinecite{Stadler06}. As shown in Fig.~\ref{fig:8}, the MO1 
changes linearly with extra partial charges. Here a correction due to the 
discontinuity in the derivative of the exchange-correlation potential with 
respect to electron density within DFT is not taken into account. The 
molecular orbitals of geometry {\bf 2} are modified upon adsorption onto 
the gold surface due to the interaction with the surface. 
Lying deeper in energy than the first occupied orbital of the 
gold slab, the MO1 is not hybridized with the gold surface. 
Thus, the shift of the MO1 in energy upon adsorption on the 
surface can be considered purely due to the charge transfer
from the surface to the molecule, assuming that the Coulomb repulsion is
small. The calculated shift of the MO1 between the whole structure
and neutral geometry {\bf 2}, 363~meV, leads to a charge transfer of 
0.28 electrons, as depicted in Fig.~\ref{fig:8}.

\subsection{Spin magnetic moment difference}

\begin{figure}
\includegraphics[angle=0, width=0.45\textwidth]{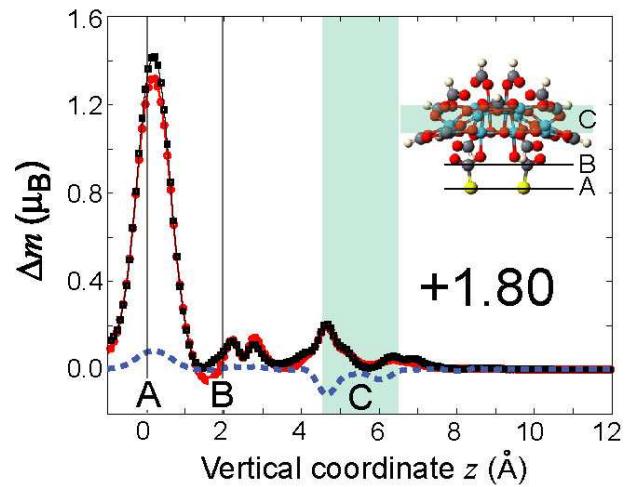}
\caption{\label{fig:9}(Color online) 
 {\it Red filled circles}: Spin magnetic moment difference
between the whole structure and geometry {\bf 2}.
The difference of the total magnetic moment is 1.80$\mu_B$
when the spin density difference is integrated from $z=-0.93$
to $z$=12~\AA.{} {\it Black empty squares}: Magnetic moment 
difference between geometry {\bf 3} and {\bf 2}.
{\it Blue dashed curve}: Magnetic moment difference 
between geometries {\bf 4} and {\bf 2}. The inset shows 
lines {\bf A} and {\bf B} and region {\bf C}.}
\end{figure}

The spin magnetic moment difference between the whole structure and geometry 
{\bf 2} is shown in Fig.~\ref{fig:9}. The increase in the total magnetic moment of 
the molecule upon adsorption amounts to 1.80 $\mu_B$ when the one-dimensional 
spin density difference is integrated from $z=-0.93$ to $z$=12~\AA{}. 
As shown in Fig.~\ref{fig:9} and Table~\ref{ta:table2}, 
the major difference arises from the sulfur atoms due to the strong bonds 
to the gold surface, while the next considerable difference occurs at 
the O anions and Mn ions close to the surface. Since only electrons 
with majority spin can be transferred to the molecule, the Mn(I) and Mn(II) 
ions close to the surface have higher magnetic moments than those for geometry 
{\bf 2}. The calculated difference of the spin magnetic moment 
corroborates the difference of the total ground-state
magnetic moment between the whole structure and geometry {\bf 2}. 
It is emphasized that this large change of the total magnetic moment 
is mainly caused by rearrangement of the magnetic moments of the S 
atoms and Mn ions due to the strong bonds to the gold surface, rather 
than by the charge transfer. Note that the total electronic charge 
must be conserved but not necessarily the total spin
magnetic moment. A spin magnetic moment of 0.20~$\mu_B$ is obtained
for the gold slab up to $z=-0.93$~\AA.{} As illustrated in Fig.~\ref{fig:9},
the magnetic moment difference between geometries {\bf 3} and {\bf 2}, 
is remarkably similar to that between the whole structure and geometry 
{\bf 2}. The two sets of the magnetic moment differences are, however, 
qualitatively dissimilar to the magnetic moment difference caused 
by 0.3 free electrons, shown by the dashed curve in Fig.~\ref{fig:9}.
The extreme similarity in the one-dimensional charge and magnetic moment 
distributions in geometry {\bf 3} and the whole structure, 
as well as the moderate level broadening in the presence of gold,
suggest that one estimate the MAB for the whole structure 
from a calculation on geometry {\bf 3}.

\subsection{Magnetic anisotropy barrier (MAB)}

\begin{table}
\caption{\label{ta:table4} Magnetic anisotropy barrier obtained
from self-consistent SOI calculations for isolated Mn$_{12}$ molecules 
(geometries {\bf 1}, {\bf 2} and {\bf 3}) and the whole structure,
in units of K.}
\begin{ruledtabular}
\begin{tabular}{ccc} 
Geometry {\bf 1}  &  Geometry {\bf 2}  &  Geometry {\bf 3}, whole structure  \\
\hline
66.7 & 66.9 & 60.7 
\end{tabular}
\end{ruledtabular}
\end{table}

The magnetic anisotropy comes from Jahn-Teller distortion around the eight 
Mn$^{3+}$ ions at the outer ring. 
It is known experimentally that for geometry {\bf 1} the second-order 
magnetic anisotropy contributes to the total MAB by 55~K and the 
fourth-order anisotropy by 10~K. \cite{BARR97}
With the total magnetic moment fixed, the MAB is computed from the change 
of the total energy upon a rotation of the spin quantization axis from 
the $z$ to the $x$ axis. The MAB calculated from a non self-consistent
SOI calculation is 57.2~K for geometry {\bf 1}, while the barrier from 
a self-consistent SOI calculation increases up to 66.7~K. 
The latter is more precise and may include higher-order magnetic anisotropy. 
This result agrees with the all-electron DFT-calculated barrier \cite{PEDE99} and 
experimental values.\cite{BARR97,HILL98} Henceforth we report only 
the MAB obtained self-consistently. The MAB with total magnetic moment 
of 18~$\mu_B$ is 66.9~K for geometry {\bf 2} so it remains the same as 
that for geometry {\bf 1}. The transverse magnetic anisotropy is calculated
to be 0.02~K, which is caused by the sulfur atoms. As discussed, 
geometry {\bf 3} is used to calculate the MAB for the whole structure.
With the total magnetic moment of 20~$\mu_B$, 
the calculated MAB for geometry {\bf 3} is 60.7~K that is 6~K lower than 
the MAB for geometries {\bf 1} and {\bf 2} (Table~\ref{ta:table4}). 
Thus, a reasonable value
for the whole structure would be also 60.7~K. The reduction
in the barrier is due to the decrease in the single-ion anisotropy for
the two Mn(II) ions close to the surface.



\section{Conclusion}
We have investigated, within DFT, the interaction between a single-molecule 
magnet Mn$_{12}$ monolayer and a Au(111) surface through thiol bonds without
long alkane chains. 
Despite a very short bond length between the sulfur-terminated Mn$_{12}$ 
molecule and the gold slab, the broadening of the molecular orbitals was 
much less than the single-electron charging energy of the molecule. 
In the ground state the sulfur-terminated Mn$_{12}$ molecule has a total
magnetic moment of 18~$\mu_B$, but its total magnetic moment increases
to 20$\mu_B$ upon adsorption onto the gold surface. 
The noticeable charge transfer from the surface to the Mn$_{12}$ 
is attributed to (i) the relative position of the gold Fermi level to the LUMO
and (ii) the long tail in the electronic cloud at the boundary of the gold 
surface.
The self-consistent SOI calculation suggested a decrease in the MAB by 9\% and 
a significant transverse magnetic anisotropy for the whole structure compared
to the isolated Mn$_{12}$. Although a relaxation of the whole structure may bring
additional transverse magnetic anisotropy caused by minor symmetry breaking 
of the Mn ions close to the surface, it is unlikely that the
molecules are completely collapsed upon adsorption. The analysis and results 
given in this study are not limited to the Mn$_{12}$ molecules and
may be applicable to systems where other types of magnetic molecules or SMMs 
are adsorbed onto nonmagnetic surfaces or bridged between nonmagnetic electrodes.
 
\begin{acknowledgments}
M.C.A. and K.P. were supported by the Jeffress Memorial Trust Funds. 
Computational support was provided by the SGI Altix Linux Supercluster
at the National Center for Supercomputing Applications under DMR060009N 
and by Virginia Tech Linux clusters and Advanced Research Computing.
\end{acknowledgments}

\end{document}